\documentclass[preprint,aps]{revtex4}

\usepackage{epsfig}

\begin{document}

\title{Impact of a global quadratic potential on galactic rotation curves}

\author{Philip~D.~Mannheim and James~G.~O'Brien}

\affiliation{Department of Physics\\ University of Connecticut\\ Storrs, CT
06269, USA
\\ {\tt 
philip.mannheim@uconn.edu,obrien@phys.uconn.edu}}

\date{November 21, 2010}

\begin{abstract}
We have made a conformal gravity fit to an available sample of 110 spiral galaxies, and report here on the 20 of those galaxies whose rotation curve data points extend the furthest from galactic centers.  We identify the impact on the 20 galaxy data set of a universal de Sitter-like potential term $V(r)=-\kappa c^2r^2/2$ that is induced by inhomogeneities in the cosmic background. This quadratic term accompanies a universal linear potential term $V(r)=\gamma_0c^2r/2$ that is associated with the cosmic background itself. We find that when these two potential terms are taken in conjunction with the contribution generated by the local luminous matter within the galaxies, the conformal theory is able to account for the rotation curve systematics that is observed in the entire 110 galaxy sample, without the need for any dark matter whatsoever. With the two universal coefficients being found to be of global magnitude, viz.  $\kappa =9.54\times 10^{-54}~{\rm cm}^{-2}$ and $\gamma_0=3.06\times 10^{-30}{\rm cm}^{-1}$, our study suggests that invoking the presence of dark matter may be nothing more than an attempt to describe global effects in purely local galactic terms. With the quadratic potential term having negative sign, galaxies are only able to support bound orbits up to distances of order $\gamma_0/\kappa = 3.21\times 10^{23}~{\rm cm}$, with global physics thus imposing a natural limit on the size of galaxies.
\end{abstract}

\maketitle

\section{Introduction}
\label{s1}

At the present time it is widely believed that on scales much larger than solar-system-sized ones astrophysical and cosmological phenomena are controlled by dark matter and dark energy, with luminous matter being only a minor contributor. However, given the lack to date of either direct detection of dark matter particles or of a solution to the cosmological constant problem, a few authors (see e.g. \cite{Mannheim2006} for a recent review) have ventured to suggest that the standard dark matter/dark energy picture may be incorrect, and that one instead needs to modify the standard Newton-Einstein gravitational theory that leads to that picture in the first place. In this paper we study one specific  alternative to Einstein gravity that has been advanced, namely conformal gravity. We report here on the results of a conformal gravity study of the instructive 20 largest of a full sample of 110 galaxies, all of whose rotation curves we have been able to fit without the need for any dark matter at all. 

In seeking an alternative to Einstein gravity that is to address both the dark matter and dark energy problems, our strategy is to seek some alternate, equally metric-based theory of gravity that possesses all of the general coordinate invariance and equivalence principle structure of Einstein gravity, that yields a geometry that is described by the Ricci-flat Schwarzschild metric on solar-system-sized distance scales while departing from it on larger ones, and that has a symmetry that can control the cosmological constant $\Lambda$. All of these criteria are met in the conformal gravity theory (see e.g. \cite{Mannheim2006}) that was first developed by Weyl. Specifically, as well as coordinate invariance, in addition one requires that the action be left invariant under local  conformal transformations of the form $g_{\mu\nu}(x)\rightarrow e^{2\alpha(x)}g_{\mu\nu}(x)$ with arbitrary local phase $\alpha(x)$.  Given this requirement, the gravitational action is then uniquely prescribed to be of the form $I_{\rm W}=-\alpha_g\int d^4x (-g)^{1/2}C_{\lambda\mu\nu\kappa}C^{\lambda\mu\nu\kappa}=-2\alpha_g\int d^4x (-g)^{1/2}\left[R_{\mu\kappa}R^{\mu\kappa}-(1/3) (R^{\alpha}_{\phantom{\alpha}\alpha})^2\right]$ where $C^{\lambda\mu\nu\kappa}$  is the conformal Weyl tensor and  $\alpha_g$ is a dimensionless gravitational coupling constant.

With the conformal symmetry forbidding the presence of any fundamental $\Lambda$ term in  $I_{\rm W}$, conformal gravity has a control on $\Lambda$ that is not possessed by Einstein gravity; and through this control conformal gravity is then able to solve the cosmological constant problem \cite{Mannheim2009,Mannheim2010a}. In addition, the conformal gravity equations of motion are given by \cite{Mannheim2006}
\begin{eqnarray}
4\alpha_g W^{\mu\nu}&=&4\alpha_g\left[
2C^{\mu\lambda\nu\kappa}_
{\phantom{\mu\lambda\nu\kappa};\lambda;\kappa}-
C^{\mu\lambda\nu\kappa}R_{\lambda\kappa}\right]=4\alpha_g\left[W^{\mu
\nu}_{(2)}-\frac{1}{3}W^{\mu\nu}_{(1)}\right]=T^{\mu\nu},
\nonumber\\
W^{\mu \nu}_{(1)}&=& 
2g^{\mu\nu}(R^{\alpha}_{\phantom{\alpha}\alpha})          
^{;\beta}_{\phantom{;\beta};\beta}                                              
-2(R^{\alpha}_{\phantom{\alpha}\alpha})^{;\mu;\nu}                           
-2 R^{\alpha}_{\phantom{\alpha}\alpha}
R^{\mu\nu}                              
+\frac{1}{2}g^{\mu\nu}(R^{\alpha}_{\phantom{\alpha}\alpha})^2,
\nonumber\\
W^{\mu \nu}_{(2)}&=&
\frac{1}{2}g^{\mu\nu}(R^{\alpha}_{\phantom{\alpha}\alpha})   
^{;\beta}_{\phantom{;\beta};\beta}+
R^{\mu\nu;\beta}_{\phantom{\mu\nu;\beta};\beta}                     
 -R^{\mu\beta;\nu}_{\phantom{\mu\beta;\nu};\beta}                        
-R^{\nu \beta;\mu}_{\phantom{\nu \beta;\mu};\beta}                          
 - 2R^{\mu\beta}R^{\nu}_{\phantom{\nu}\beta}                                    
+\frac{1}{2}g^{\mu\nu}R_{\alpha\beta}R^{\alpha\beta},
\label{108}
\end{eqnarray}                                 
with Schwarzschild  thus being a vacuum solution to conformal gravity, just as required \cite{footnote1}.

\section{Universal Potentials from the Rest of the Universe}
\label{s2}

Given its structure, $W^{\mu\nu}$ could potentially vanish even if the geometry is  not Ricci flat, and the conformal theory could thus have non-Schwarzschild solutions as well. To identify any such solutions, Mannheim and Kazanas solved for the metric in a vacuum region exterior to a static, spherically symmetric source of radius $r_0$, to find \cite{Mannheim1989} that the exact, all-order line element is given by $ds^2=-B(r)dt^2+dr^2/B(r)+r^2d\Omega_2$, with
\begin{equation}
B(r>r_0)=1-\frac{2\beta}{r}+\gamma r -kr^2.
\label{E1}
\end{equation}
In the $\gamma r$ and $-kr^2$ terms we see that the conformal gravity metric  departs from the $B(r>r_0)=1-2\beta/r$ Schwarzschild metric only at large $r$, just as we want.

In seeking to relate the various constants in (\ref{E1}) to properties of the energy-momentum tensor $T_{\mu\nu}$ of the source, Mannheim and Kazanas found \cite{Mannheim1994} that in the static, spherically symmetric case the quantity $(3/B(r))\left(W^0_{{\phantom 0} 0} - W^r_{{\phantom r} r}\right)$ evaluates exactly to $\nabla^4B(r)$; and that, in terms of the general source function $f(r)=(3/4\alpha_gB(r))(T^{0}_{\phantom{0}0}-T^{r}_{\phantom{r}r})$, the exact fourth-order equation of motion of the conformal theory reduced to the remarkably simple 
\begin{equation}
\nabla^4B=\left[\frac{d^4}{dr^4}+\frac{4}{r}\frac{d^3}{dr^3}\right]B(r)=f(r),
\label{E2}
\end{equation}
without any approximation whatsoever. Since $\nabla^4(r^2)$ vanishes identically everywhere while $\nabla^4(1/r)$ and $\nabla^4(r)$ evaluate to delta functions and their derivatives, we see that of the constants given in (\ref{E1}), only $\beta$ and $\gamma$, but not $k$, can be associated with properties of a local source of radius $r_0$; with the matching of the interior and exterior metrics then yielding \cite{Mannheim1994} 
\begin{equation}
\gamma= -\frac{1}{2}\int_0^{r_0}dr^{\prime}r^{\prime 2}f(r^{\prime}),\qquad 
2\beta=\frac{1}{6}\int_0^{r_0}dr^{\prime}r^{\prime 4}f(r^{\prime}).
\label{E3}
\end{equation}
Thus despite  the presence of the $-kr^2$ term in the exterior vacuum solution in (\ref{E1}), the above analysis provides no specific basis for considering it further, as it is associated with the trivial solution to $\nabla^4B=0$, to thereby be devoid of dynamical content.

In conformal gravity a local gravitational source generates a gravitational potential
\begin{equation}
V^{*}(r)=-\frac{\beta^{*}c^2}{r}+\frac{\gamma^{*} c^2 r}{2}
\label{E4}
\end{equation}
per unit solar mass, with $\beta^{*}$ being given by the familiar $M_{\odot}G/c^2=1.48\times 10^{5}$ cm, and with the numerical value of the solar $\gamma^{*}$ needing to be determined by data fitting. In the theory the visible local material in a given galaxy would generate a net local gravitational potential $V_{\rm  LOC}(r)$ given by integrating $V^{*}(r)$ over the visible galactic mass distribution. In disk galaxies luminous matter is typically distributed with a surface brightness $\Sigma (R)=\Sigma_0e^{-R/R_0}$ with scale length $R_0$ and total luminosity $L=2\pi \Sigma_0R_0^2$, with most of the surface brightness being contained in the $R \leq 4R_0$ or so optical disk region. For a galactic mass to light ratio $M/L$, one can define the total number of solar mass units $N^{*}$ in the galaxy via $(M/L)L=M=N^{*}M_{\odot}$. Then, on integrating $V^{*}(r)$ over this visible matter distribution,  one obtains \cite{Mannheim2006} the net local luminous contribution  

\begin{eqnarray}
\frac{v_{{\rm LOC}}^2}{R}&=&
\frac{N^*\beta^*c^2 R}{2R_0^3}\left[I_0\left(\frac{R}{2R_0}
\right)K_0\left(\frac{R}{2R_0}\right)-
I_1\left(\frac{R}{2R_0}\right)
K_1\left(\frac{R}{2R_0}\right)\right]
\nonumber \\
&&+\frac{N^*\gamma^* c^2R}{2R_0}I_1\left(\frac{R}{2R_0}\right)
K_1\left(\frac{R}{2R_0}\right)
\label{E5}
\end{eqnarray} 
for the centripetal accelerations of particles orbiting in the plane of the galactic disk. 

However, unlike the situation that obtains in standard second-order gravity, one cannot simply use (\ref{E5}) as is to fit galactic rotation curve data, as there are two additional global effects coming from the rest of the material in the universe that need to be taken into consideration as well, one associated with the homogeneous cosmological background and the other with the inhomogeneities in it. As regards first the effect of inhomogeneities, we recall for the standard second order Poisson equation $\nabla^2\phi(r)=g(r)$, the force associated with a general static, spherically symmetric source $g(r)$  is given by 
\begin{equation}
\frac{d\phi(r)}{dr}= \frac{1}{r^2}\int_0^r
dr^{\prime}r^{\prime 2}g(r^{\prime}).
\label{E8}
\end{equation}                                 
As such, the import of (\ref{E8}) is that even though $g(r)$ could continue globally all the way to infinity, the force at any radial point $r$ is determined only by the material in the local $0< r^{\prime}< r$ region. In this sense Newtonian gravity is local, since to explain a gravitational effect in some local region one only needs to consider the material in that region. Thus in Newtonian gravity, if one wishes to explain the behavior of galactic rotation curves through the use of dark matter, one must locate the dark matter where the problem is and not elsewhere. Since the discrepancy problem in galaxies occurs primarily in the region beyond the optical disk, one must thus locate galactic dark matter in precisely the region in galaxies where there is little or no visible matter.

However, this local character to Newtonian gravity is not a generic property of any gravitational potential. In particular for the fourth-order Poisson equation $\nabla^4\phi(r)=h(r)=f(r)c^2/2$ of interest to conformal gravity, the potential and the force evaluate to
\begin{eqnarray}
\phi(r)&= &-\frac{r}{2}\int_0^r
dr^{\prime}r^{\prime 2}h(r^{\prime})
-\frac{1}{6r}\int_0^r
dr^{\prime}r^{\prime 4}h(r^{\prime})
-\frac{1}{2}\int_r^{\infty}
dr^{\prime}r^{\prime 3}h(r^{\prime})
-\frac{r^2}{6}\int_r^{\infty}
dr^{\prime}r^{\prime }h(r^{\prime}),
\nonumber\\
\frac{d\phi(r)}{dr}&=& -\frac{1}{2}\int_0^r
dr^{\prime}r^{\prime 2}h(r^{\prime})
+\frac{1}{6r^2}\int_0^r
dr^{\prime}r^{\prime 4}h(r^{\prime})
-\frac{r}{3}\int_r^{\infty}
dr^{\prime}r^{\prime }h(r^{\prime}),
\label{E10}
\end{eqnarray}                                 
so that this time we do find a global contribution to the force coming from material that is beyond the radial point of interest. Hence in conformal gravity one cannot ignore the rest of the universe, with a test particle in orbit in a galaxy being able to sample both the local field due to the matter in the galaxy and the global field due to the rest of the universe.

In the presence of inhomogeneities $W^{\mu\nu}$ does not vanish, as the very presence of a localized source prevents a geometry from being conformal to flat, with inhomogeneities in the universe thus leading to integrals in (\ref{E10}) that can extend to very large distances. However, this is not the only global effect that we need to take into consideration, as one can also add on to (\ref{E10}) any terms that would cause $W^{\mu\nu}$ to vanish, provided they make it do so non-trivially.  Since the cosmological Robertson-Walker (RW) metric is homogeneous and isotropic, it is conformal to flat, and thus its geometry obeys $W^{\mu\nu}=0$. For the cosmological background the vanishing of $W^{\mu\nu}$ entails that conformal cosmology be described by $T^{\mu\nu}=0$. As discussed in \cite{Mannheim2006} the equation $T^{\mu\nu}=0$ can be satisfied non-trivially, and leads to a topologically open RW cosmology, with its contribution to $W^{\mu\nu}$ then  vanishing non-trivially, just as desired. 

Since cosmology is written in comoving Hubble flow coordinates while rotation curves are measured in galactic rest frames, to ascertain the impact of cosmology on rotation curves one needs to transform the RW metric to static coordinates. As noted in \cite{Mannheim1989}, the transformation
\begin{equation}
\rho=\frac{4r}{2(1+\gamma_0r-k r^2)^{1/2}+2 +\gamma_0 r},\qquad \tau=\int dt R(t)
\label{E11}
\end{equation}                                 
effects the metric transformation
\begin{eqnarray}
&&-(1+\gamma_0r-k r^2)c^2dt^2+\frac{dr^2}{(1+\gamma_0r-k r^2)}+r^2d\Omega_2=
\nonumber \\
&&\frac{1}{R^2(\tau)}\frac{[1-\rho^2(\gamma_0^2/16+k /4)]^2}
{[(1-\gamma_0\rho/4)^2+k\rho^2/4]^2}
\left[-c^2d\tau^2+\frac{R^2(\tau)}{[1-\rho^2(\gamma_0^2/16+k /4)]^2}
\left(d\rho^2+\rho^2d\Omega_2\right)\right].
\label{E12}
\end{eqnarray} 
Since an RW geometry is conformally flat and since it remains so under a conformal transformation, we see that when written in a static coordinate system, a comoving conformal cosmology with 3-space spatial curvature $K$ looks just like a static metric with universal linear and quadratic terms with coefficients that obey $K=-\gamma_0^2/4-k $.  However, since there was only one spatial scale in the RW metric (viz. $K$), its decomposition into two static coordinate system scales ($\gamma_0$ and $k$) was artificial, and so in \cite{Mannheim1997} the $k $ term was dropped. Then, without the $k$ term, we see that in the rest frame of a comoving galaxy (i.e. one with no peculiar velocity with respect to the Hubble flow),  a topologically open  RW cosmology would look just like a universal linear potential with cosmological strength $\gamma_0/2=(-K)^{1/2}$. 

In the conformal theory then we recognize not one but two linear potential terms, a local $N^*\gamma^*$ dependent one associated with the matter within a galaxy and a global cosmological one $\gamma_0c^2r/2$ associated with cosmological background. Thus in the weak gravity limit one can add the two potentials and replace (\ref{E5}) by \cite{Mannheim1997} 
\begin{equation}
\frac{v^2_{\rm TOT}}{R}=\frac{v^2_{\rm LOC}}{R}+\frac{\gamma_0 c^2}{2}.
\label{E16}
\end{equation}                                 
In \cite{Mannheim1997}  (\ref{E16}) was used to fit the galactic rotation curve data of a sample of 11 galaxies  (of which only NGC 2841 and NGC 3198 are in the sample considered here), and good fits were found, with the two universal linear potential parameters being fixed to the values 
\begin{equation}
\gamma^*=5.42\times 10^{-41} {\rm cm}^{-1},\qquad \gamma_0=3.06\times
10^{-30} {\rm cm}^{-1}.
\label{E18}
\end{equation} 
The value obtained for $\gamma^*$ entails that the linear potential of the Sun is so small that there are no modifications to standard solar system phenomenology, with the values obtained for $N^*\gamma^*$ and $\gamma_0$ being such that one has to go to galactic scales before their effects can become as big as the Newtonian contribution. The value obtained for $\gamma_0$ shows that it is indeed of cosmological magnitude. In the fitting to the 110 galaxy sample (\ref{E18}) does not change.

However, as we had noted above, there is a contribution due to inhomogeneities in the cosmic background that we need to include too. These inhomogeneities would typically be clusters and superclusters and would be associated with distance scales between 1 Mpc and 100 Mpc or so. Without knowing anything other than that about them, we see from (\ref{E10}) that  for calculating potentials at galactic distance scales (viz. scales much less than cluster scales) the inhomogeneities would contribute constant and quadratic terms multiplied by integrals that are evaluated between end points that do not depend on the galaxy of interest, to thus be constants.  Thus, again up to peculiar velocity effects, we augment (\ref{E16}) to 
\begin{equation}
\frac{v^2_{\rm TOT}}{R}=\frac{v^2_{\rm LOC}}{R}+\frac{\gamma_0 c^2}{2}-\kappa c^2R,
\label{E20}
\end{equation}                                 
with asymptotic limit 
\begin{equation}
\frac{v_{{\rm TOT}}^2}{R} \rightarrow \frac{N^*\beta^*c^2}{R^2}+
\frac{N^*\gamma^*c^2}{2}+\frac{\gamma_0c^2}{2}-\kappa c^2R.
\label{E21}
\end{equation} 
It is thus (\ref{E20}) with its universal $\kappa $ that we must use for fitting galactic rotation curves, and in making such fits the only parameter that can vary from one galaxy to the next is the galactic disk mass to light ratio as embodied in $N^*$. Our fits are thus highly constrained, one parameter per galaxy, fits (the fits also include the effect of HI gas, but for the gas the mass is known), with everything else being universal, and no dark matter being assumed.

\section{Data Fitting}
\label{s3}

We recall that in \cite{Mannheim1997} successful rotation curve fitting to an 11 galaxy sample was obtained using (\ref{E16}), and one would thus initially anticipate that even if the $-\kappa c^2R$ term were to be present in principle, in practice it would be too small to have any effect. However, the sample we have studied now is altogether larger (110 galaxies) and it contains some very instructive galaxies whose data points extend to larger distances from galactic centers than had been the case for the 11 galaxy sample studied in \cite{Mannheim1997}. It is through fitting these highly extended galaxies that we are able to uncover a role for the $-\kappa c^2R$ term and extract a value for $\kappa $ given by $\kappa =9.54\times 10^{-54}~{\rm cm}^{-2}$. And in the fitting to the full 110 galaxy sample to be reported elsewhere \cite{Mannheim2010b} (a varied sample of galaxies that includes high (HSB) and low (LSB) surface brightness galaxies and dwarfs) we are able to confirm that even with this now fixed value for  $\kappa $, (\ref{E20}) fully accounts for the data . With $\kappa $ being found to be of order $1/(100~{\rm Mpc})^2$, it is indeed an inhomogeneous rather than a Hubble distance scale.

In Fig. (1) we present our fits to the 20 galaxy sample with the relevant parameters being listed in  Table (1). In Fig. (1) the rotational velocities and errors  (in ${\rm km}~{\rm sec}^{-1}$) are plotted as a function of radial distance (in ${\rm kpc}$). For each galaxy
we exhibit the contribution due to the luminous Newtonian term alone (dashed curve), the 
contribution from the two linear terms alone (dot dashed curve), the contribution from the two 
linear terms and the quadratic terms combined (dotted curve), with the full curve showing the total contribution. As we see, without any need for dark matter, our fitting captures the essence of the data. Because the data go out to much further distances than had been the case for the sample studied in \cite{Mannheim1997}, the data are now sensitive to the rise in velocity associated with the linear potential terms, and it is here that the quadratic term acts to actually arrest  the rise altogether (dotted curve) and cause all rotation velocities to ultimately fall. Moreover, since $v^2$ cannot be negative, beyond a distance $R$ of order $\gamma_0/\kappa  =3.21\times 10^{23}~{\rm cm}$ or so there could no longer be any bound galactic orbits, with galaxies thus having a natural way of terminating, and with global physics thus imposing a natural limit on the size of galaxies. To illustrate this we plot the rotation velocity curve for UGC 128 over an extended range. The fits presented here and in \cite{Mannheim2010b} are noteworthy since the universal $\gamma_0$ and $\kappa$ terms have no dependence on individual galactic  properties whatsoever and yet have to work in every single case.

 It is important to appreciate that the fits provided by conformal gravity (and likewise by other  alternate theories such as the MOND \cite{Milgrom1983} and MSTG \cite{Brownstein2006} theories are predictions. Specifically, for all these theories the only input one needs is the optical data, and the only free parameter is the $M/L$ ratio for each given galaxy, with rotation velocities then being determined \cite{footnote4}. As Table (1) shows,  by and large the $M/L$ ratios that are found are all of order the solar $M_{\odot}/L_{\odot}$ ratio, just as one would want. It is important to emphasize that the fits are predictions since dark matter fitting to galactic data works very differently. There one first needs to know the velocities so that one can then ascertain the needed amount of dark matter, i.e. in its current formulation dark matter is only a parametrization or postdiction of the velocity discrepancies that are observed and is not a prediction of them. Dark matter theory has yet to develop to the point where one is able to predict rotation velocities given a knowledge of the luminous distribution alone. Thus dark matter theories, and in particular those theories that produce dark matter halos in the early universe, are currently unable to make an a priori determination as to which halo is to go with which particular luminous matter distribution, and need to fine-tune halo parameters to luminous parameters galaxy by galaxy. No such shortcoming appears in conformal gravity, and if standard gravity is to be the correct  description of gravity, then a universal formula akin to the one given in (\ref{E20}) would need to be derived by dark matter theory. However, since our study establishes that global physics does indeed influence local galactic motions, the invoking of dark matter in galaxies could potentially be nothing more than an attempt to describe global effects in purely local galactic terms. 
We would like to thank Dr.~J.~R.~Brownstein, Dr.~W.~J.~G.~ de Blok, Dr.~J.~W.~Moffat, and Dr.~ S.~S.~McGaugh for helpful communications, and especially for providing their galactic data bases. We are particularly indebted to Dr.~McGaugh for having alerted us to the fact that a linear potential would lead to an overshoot in UGC 128.

{}

\begin{table}
\caption{Properties of the 20 Large Galaxy Sample}
\centering
\begin{tabular}{l c c c c c c c c c } 
\hline\hline
 \phantom{00}Galaxy  & \phantom{0}Type \phantom{0}&Distance  & $L_{\rm B}$ & $R_0$  & $R_{\rm last} $ &  $M_{\rm HI} $ & $M_{\rm disk}$ &  $ 
(M/L) _{\rm stars}$ & $(v^2 / c^2 R)_{\rm last}$ \\  
& &  (Mpc)  &  $(10^{10}{\rm L}_{\odot})$&(kpc) & (kpc) & {$(10^{10} M_\odot)$} & {$(10^{10}
M_\odot)$} & ({$M_\odot L_\odot^{-1}$}) & {$(10^{-30}{\rm cm}^{-1})$}  \\
\hline
NGC 3726 &HSB & 17.4&   \phantom{0}3.34    & \phantom{0}3.2&31.5&0.60&\phantom{0}3.82&1.15&3.19\\
NGC 3769 &HSB&15.5&   \phantom{0}0.68    & \phantom{0}1.5 & 32.2 &0.41 &\phantom{0}1.36 &1.99 &1.43 \\ 
NGC 4013&HSB & 18.6 & \phantom{0}2.09    &  \phantom{0}2.1 & 33.1 & 0.32 & \phantom{0}5.58 & 2.67 &3.14 \\
NGC 3521 &HSB&12.2  &   \phantom{0}4.77    &  \phantom{0}3.3   &35.3 &1.03 & \phantom{0}9.25 & 1.94 &4.21 \\ 
NGC 2683 &HSB& 10.2 &    \phantom{0}1.88   &  \phantom{0}2.4 & 36.0 &0.15 &\phantom{0}6.03 & 3.20& 2.28 \\
UGC 1230 &LSB& 54.1&   \phantom{0}0.37    &  \phantom{0}4.7 & 37.1 & 0.65 & \phantom{0}0.67 &1.82 & 0.97 \\ 
NGC 3198 &HSB&14.1&  \phantom{0}3.24     &  \phantom{0}4.0 & 38.6 &1.06 &\phantom{0}3.64 &1.12 &2.09 \\  
NGC 5371 &HSB& 35.3 &   \phantom{0}7.59    &  \phantom{0}4.4 & 41.0 & 0.89 & \phantom{0}8.52 & 1.44 & 3.98 \\ 
NGC 2998 &HSB& 59.3&   \phantom{0}5.19    &  \phantom{0}4.8 & 41.1 & 1.78 &\phantom{0}7.16 & 1.75& 3.43 \\ 
NGC 5055 &HSB& \phantom{0}9.2&  \phantom{0}3.62    & \phantom{0}2.9 & 44.4 &0.76& \phantom{0}6.04 & 1.87& 2.36\\ 
NGC 5033 &HSB& 15.3 &  \phantom{0}3.06    &  \phantom{0}7.5 & 45.6& 1.07 & \phantom{0}0.27 &3.28 & 3.16 \\ 
NGC 0801 &HSB& 63.0 &  \phantom{0}4.75     &  \phantom{0}9.5 & 46.7 & 1.39 & \phantom{0}6.93 &2.37 & 3.59 \\ 
NGC 5907 &HSB& 16.5 &   \phantom{0}5.40   &  \phantom{0}5.5 & 48.0 & 1.90 &\phantom{0}2.49 &1.89& 3.44\\ 
NGC 3992 &HSB& 25.6 &  \phantom{0}8.46     &  \phantom{0}5.7& 49.6 & 1.94 & 13.94 &  1.65 & 4.08 \\ 
NGC 2841 &HSB& 14.1 &   \phantom{0}4.74    &   \phantom{0}3.5 & 51.6 & 0.86 & 19.55 & 4.12&5.83 \\ 
UGC 0128 &LSB&  64.6 &   \phantom{0}0.60    &   \phantom{0}6.9 & 54.8 & 0.73 & \phantom{0}2.75 & 4.60 & 1.03 \\
NGC 5533 &HSB& 42.0 &  \phantom{0}3.17     &  \phantom{0}7.4 & 56.0 & 1.39 & \phantom{0}2.00 &4.14  &3.31 \\  
NGC 6674 &HSB& 42.0 &  \phantom{0}4.94     & \phantom{0}7.1 & 59.1 & 2.18 & \phantom{0}2.00&2.52 & 3.57 \\ 
UGC 6614 &LSB& 86.2 &   \phantom{0}2.11    &  \phantom{0}8.2 & 62.7 & 2.07 & \phantom{0}9.70 &  4.60 & 2.39 \\
UGC 2885 &HSB& 80.4 &  23.96     & 13.3 & 74.1 &3.98 &\phantom{0}8.47 &0.72 & 4.31 \\  

\hline
\end{tabular}
\label{table:large}
\end{table}

\begin{figure}[t]
\epsfig{file=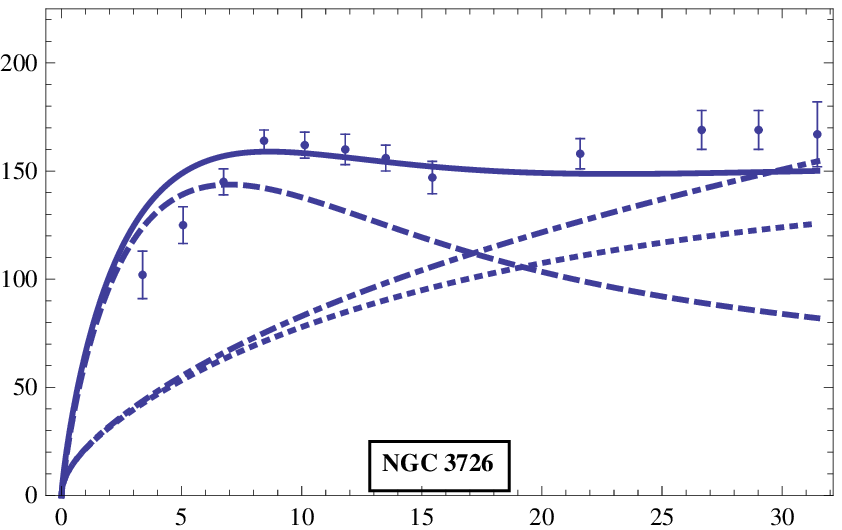,width=2.11in,height=1.2in}~~~
\epsfig{file=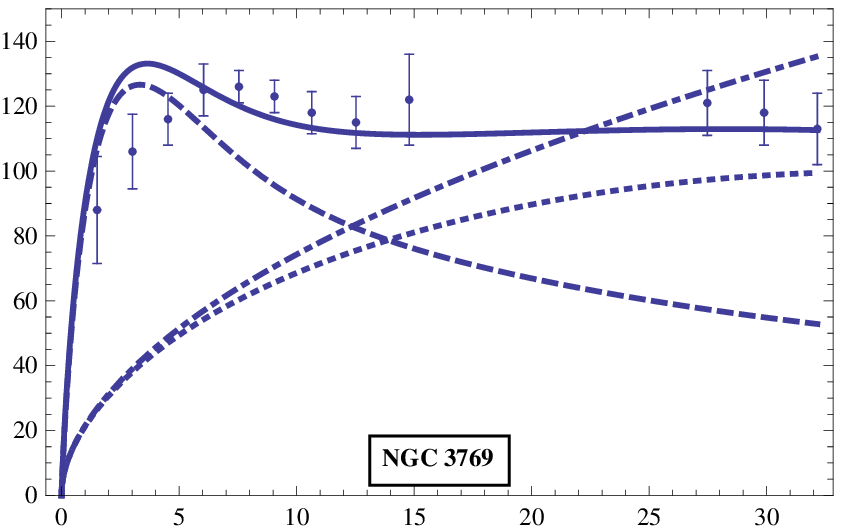,width=2.11in,height=1.2in}~~~
\epsfig{file=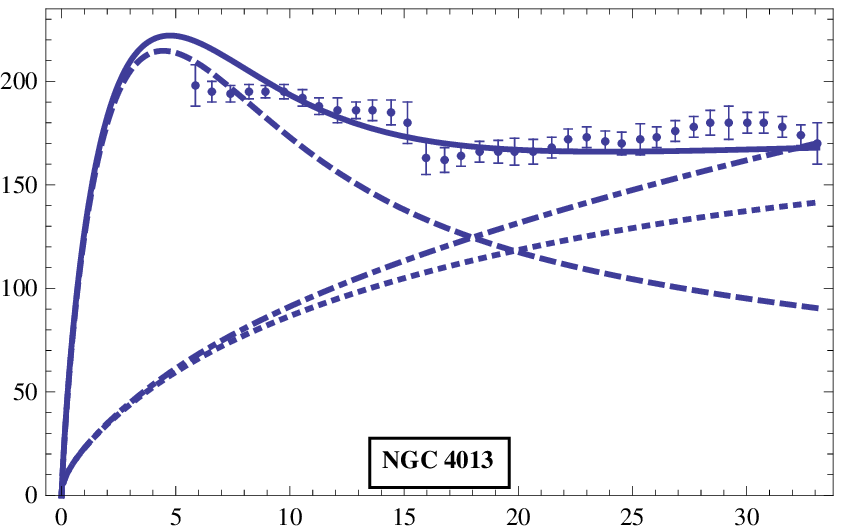,width=2.11in,height=1.2in}\\
\smallskip
\epsfig{file=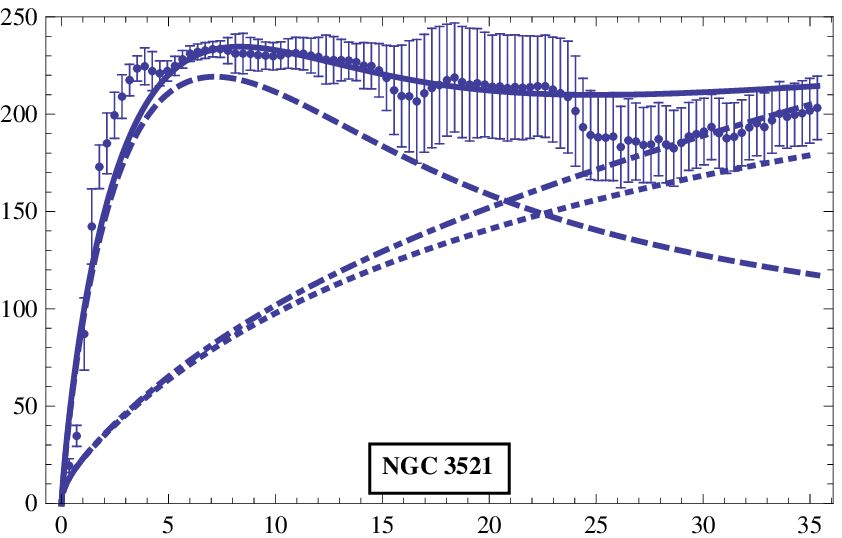,width=2.11in,height=1.2in}~~~
\epsfig{file=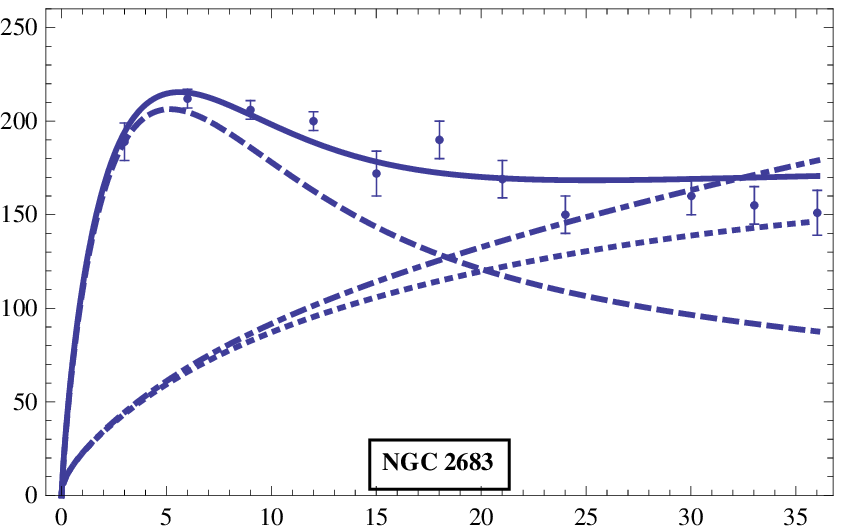,width=2.11in,height=1.2in}~~~
\epsfig{file=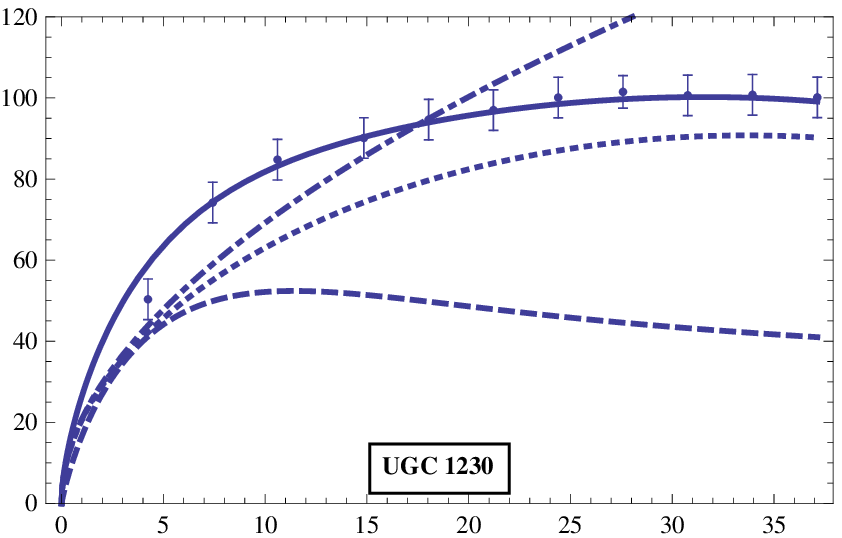,width=2.11in,height=1.2in}~~~\\
\smallskip
\epsfig{file=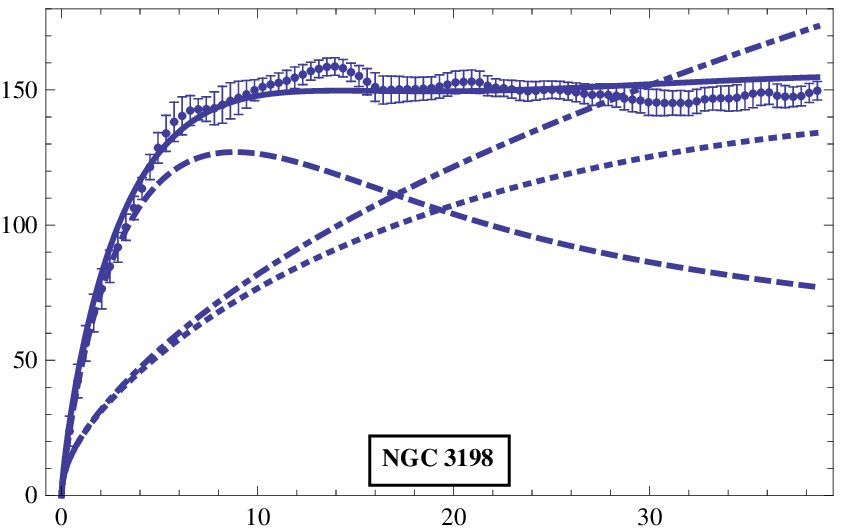,width=2.11in,height=1.2in}~~~
\epsfig{file=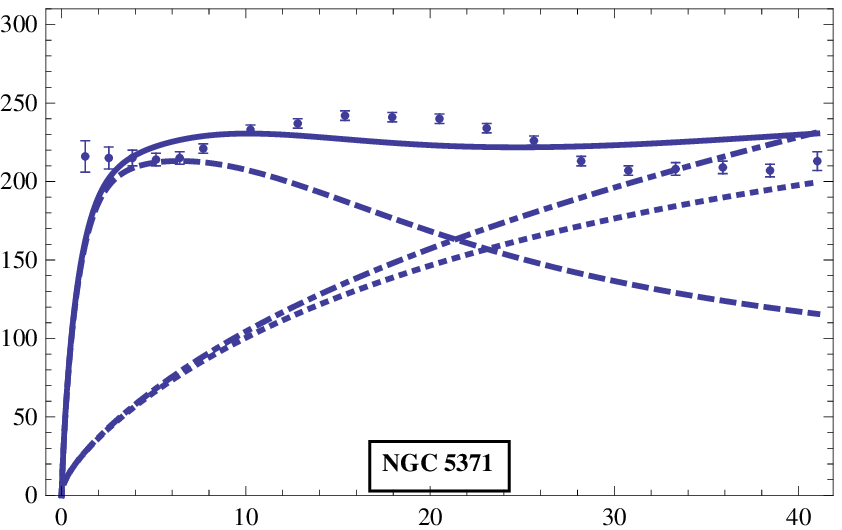,width=2.11in,height=1.2in}~~~
\epsfig{file=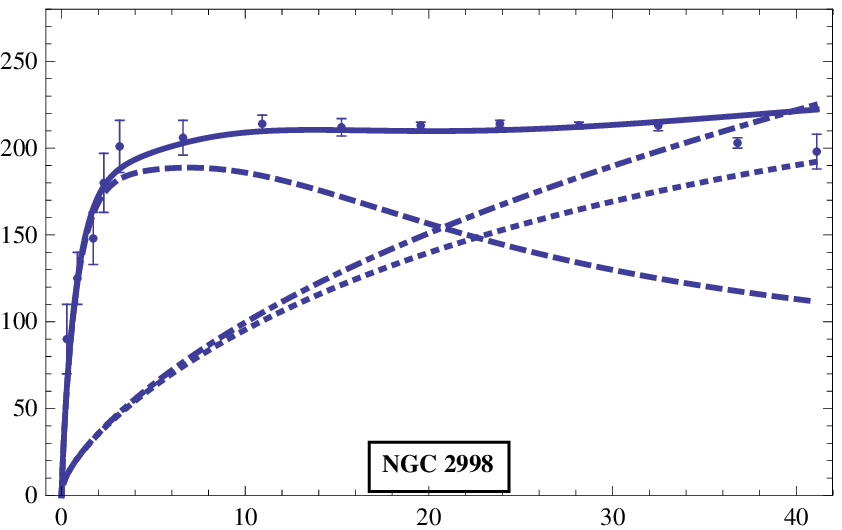,width=2.11in,height=1.2in}\\
\smallskip
\epsfig{file=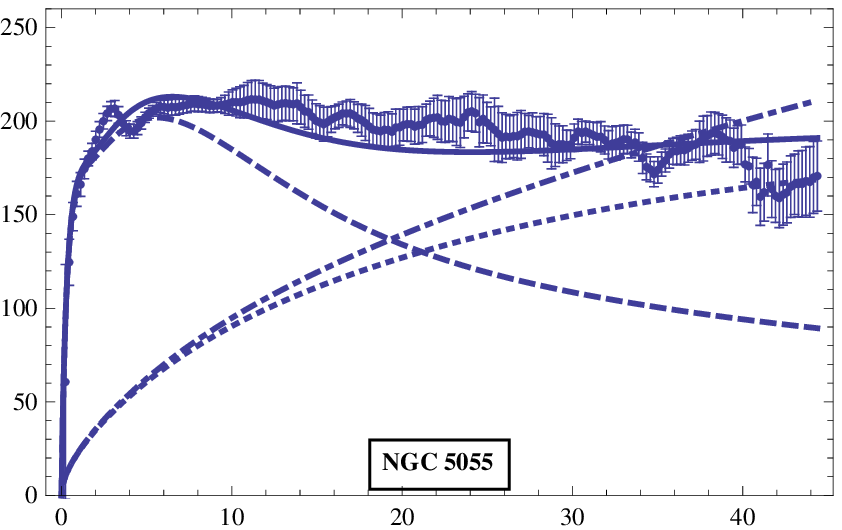,width=2.11in,height=1.2in}~~~
\epsfig{file=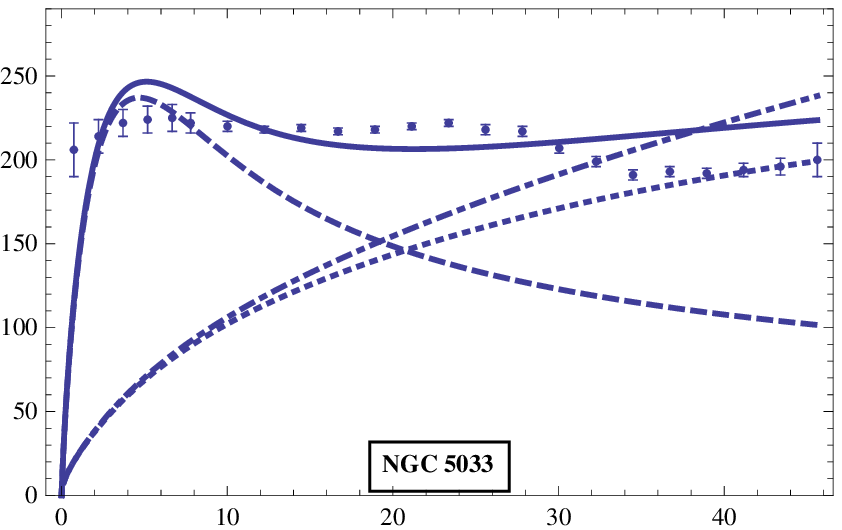,width=2.11in,height=1.2in}~~~
\epsfig{file=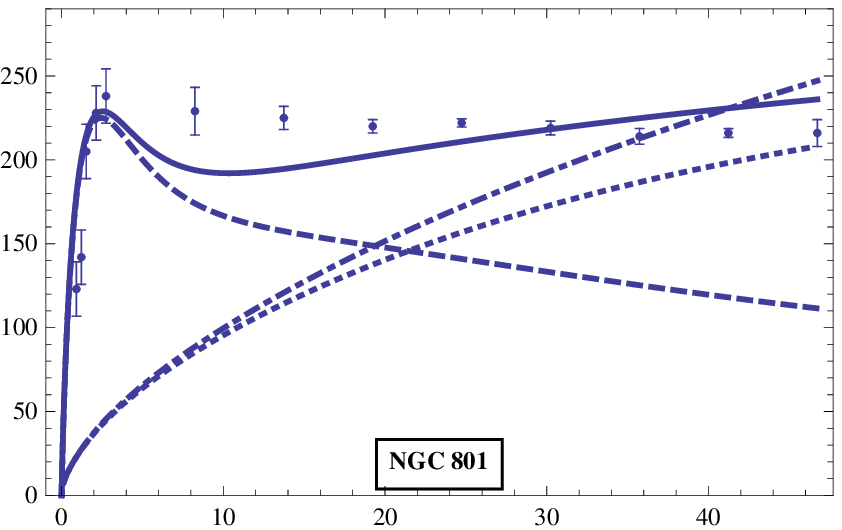,  width=2.11in,height=1.2in}\\
\smallskip
\epsfig{file=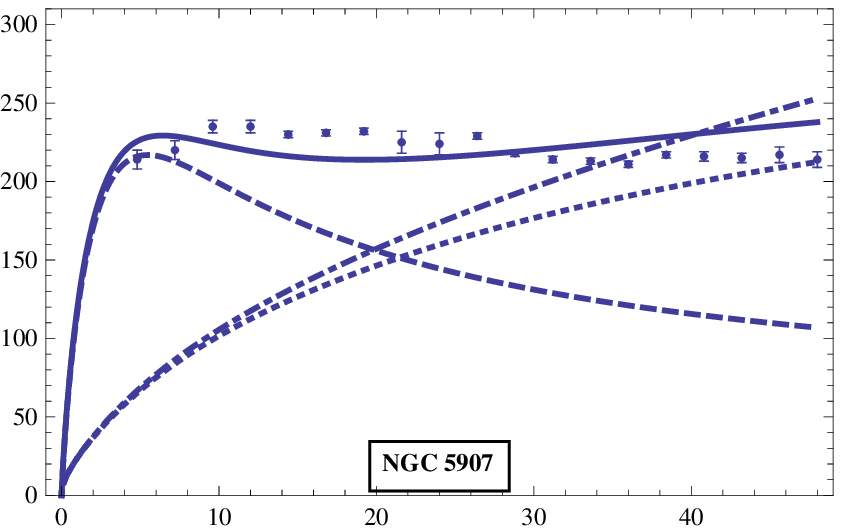,width=2.11in,height=1.2in}~~~
\epsfig{file=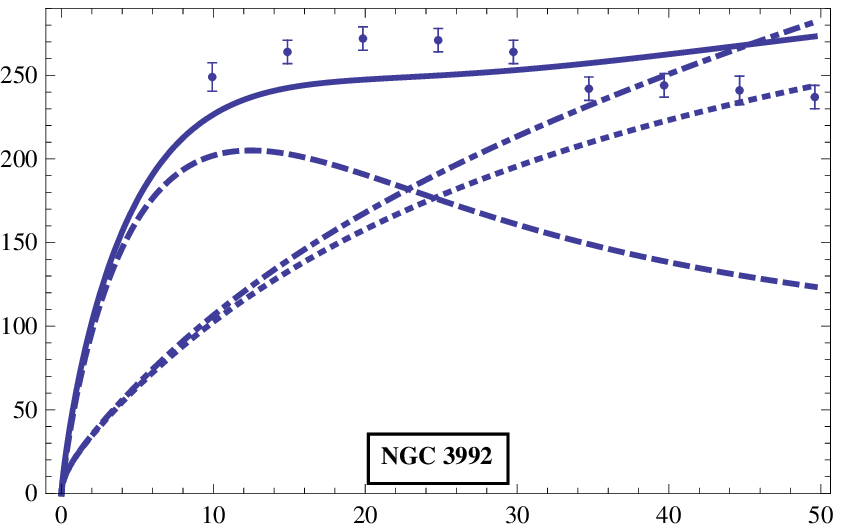,width=2.11in,height=1.2in}~~~
\epsfig{file=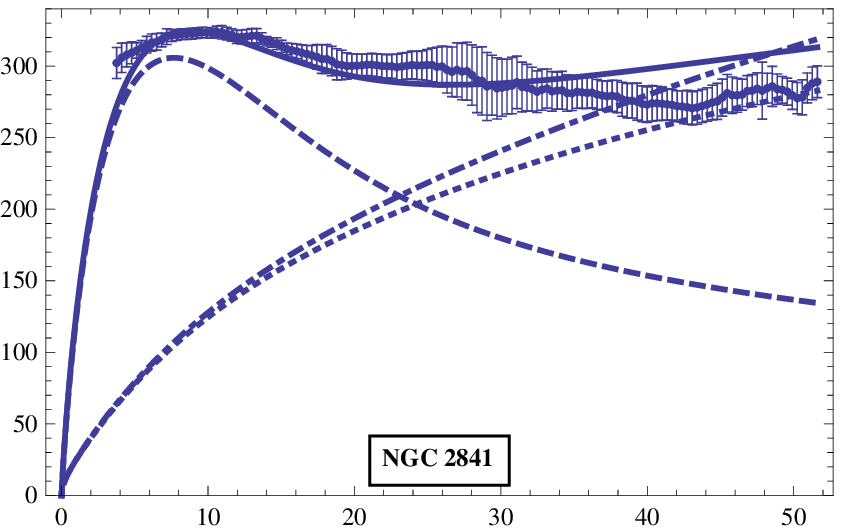,width=2.11in,height=1.2in}\\
\smallskip
\epsfig{file=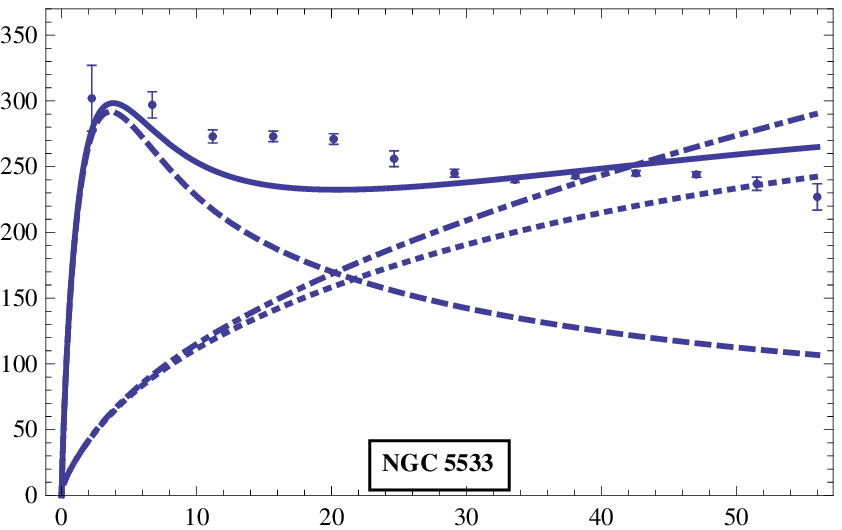,width=2.11in,height=1.2in}~~~
\epsfig{file=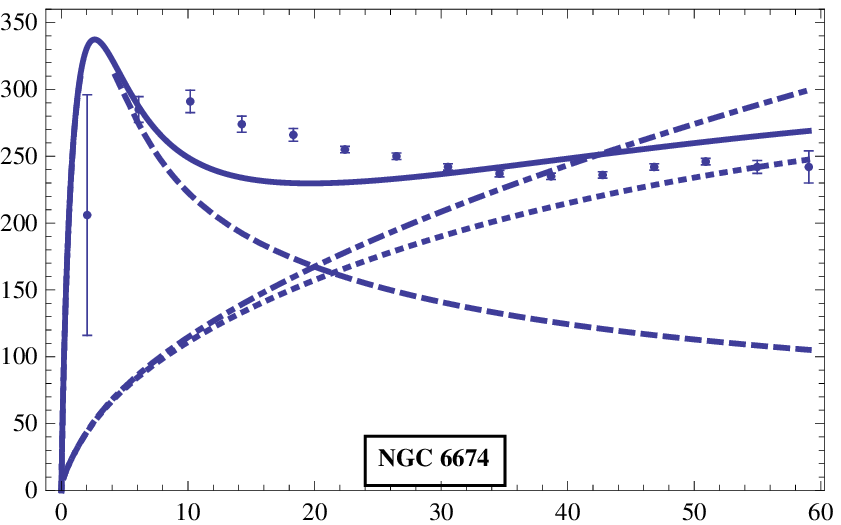,width=2.11in,height=1.2in}~~~
\epsfig{file=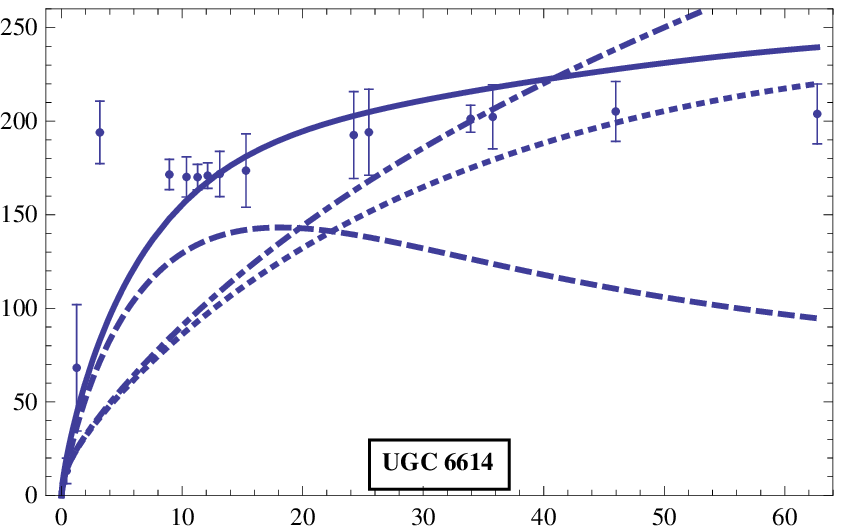,width=2.11in,height=1.2in}\\
\smallskip
\epsfig{file=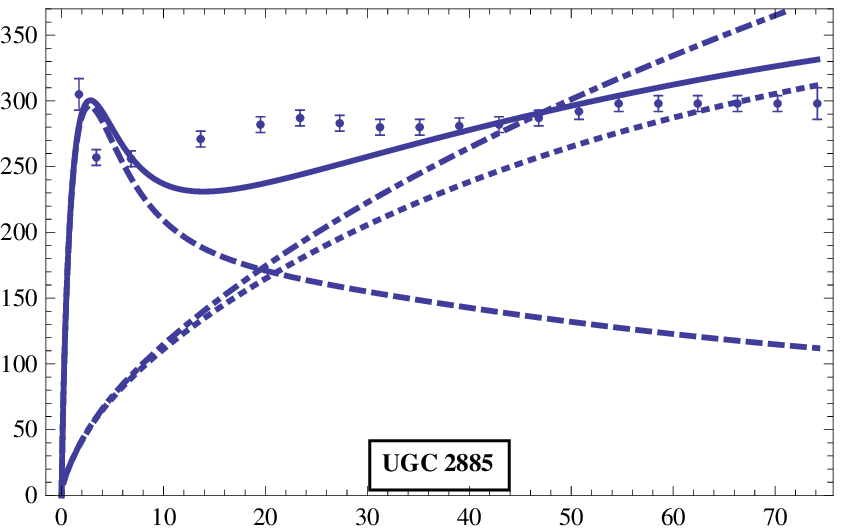,width=2.11in,height=1.2in}~~~
\epsfig{file=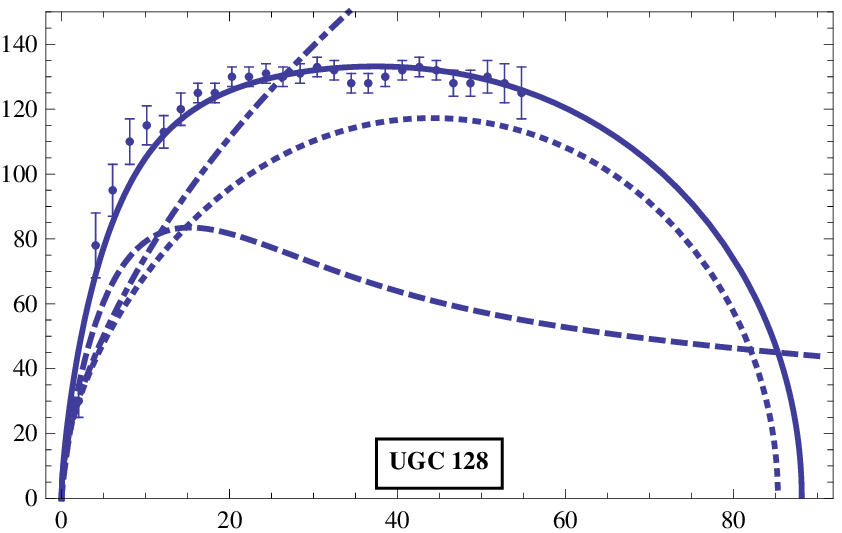,  width=3.23in,height=1.2in}\\
\medskip
\caption{Fitting to the rotational velocities of  the 20 galaxy sample}
\label{Fig. (1)}
\end{figure}


\vskip-0.65truecm

\begin{thebibliography}{}


\bibitem{Mannheim2006} P.~D.~Mannheim,~Prog.~Part.~Nucl.~Phys.~{\bf 56},~340~(2006). 

\bibitem{Mannheim2009} P.~D.~Mannheim,~{\it Comprehensive solution to the cosmological constant, zero-point energy, and quantum gravity problems}, arXiv:0909.0212 [hep-th]. Gen.~Rev.~Gravit.~in press.

\bibitem{Mannheim2010a} P.~D.~Mannheim,~{\it Intrinsically quantum-mechanical gravity and the cosmological constant problem}, May 2010 (arXiv:1005.5108 [hep-th]). 

\bibitem{footnote1} As well as being of interest here as a classical theory, in  C.~M.~Bender and P.~D.~Mannheim, Phys.~Rev.~Lett.~{\bf 100},~110402 (2008);~Phys.~Rev.~D {\bf 78},
025022 (2008) it has been shown that as a quantum theory the  fourth-order conformal gravity theory is  both consistent and unitary.


\bibitem{Mannheim1989} P.~D.~Mannheim and D.~Kazanas,~Ap.~J.~{\bf 342}, 635 (1989).

\bibitem{Mannheim1994} P.~D.~Mannheim and D.~Kazanas,~Gen.~Rel.~Gravit.~{\bf 26}, 337 (1994).

\bibitem{Mannheim1997} P.~D.~Mannheim,~Ap.~J.~{\bf 479}, 659 (1997).


\bibitem{Mannheim2010b} P.~D.~Mannheim and J.~G.~O'Brien,~{\it Fitting galactic rotation curves with conformal gravity and a global quadratic potential}, (arXiv:1011.3495 [astro-ph.CO]).  In this paper we give full references to the data sets we use. In the fits we have taken into account bulges in NGC 5371, NGC 2998, NGC 5055, NGC 5033, NGC 801, NGC 5907, NGC 5533, NGC 6674 and UGC 2885, but not allowed for a bar in NGC 6674 or lopsidedness of the bulge in NGC 5533. While the inner region rotation curve is sensitive to a disk-bulge-bar decomposition of the luminosity, only the net luminous mass appears in the asymptotic (\ref{E21}), with the outer region cancellation effected by the $-\kappa c^2R$ term not depending on how much of the net $N^*$ is disk, bulge or bar.


\bibitem{Milgrom1983} M.~Milgrom,~Ap.~J.~ {\bf 270}, 365, 371, 384 (1983).

\bibitem{Brownstein2006} J.~R.~Brownstein and  J.~W.~Moffat,~Ap.~J.~{\bf 636}, 721 (2006).

\bibitem{footnote4}  That these various alternate theories all work is because they each possess an underlying universal structure, with the last column in Table (1) and the data presented in
\cite{Mannheim2010b} indicating that in each galaxy the centripetal accelerations measured at the last data points possess it too. For the moment such universality is not explained by dark matter theory.



\end{thebibliography}
\end{document}